\documentstyle[graphicx,12pt]{article}
\begin{document}

\small
\hoffset=-1truecm
\voffset=-2truecm
\title{\bf The further analytical discussions on the $U(1)$ gauged Q-balls
with $N$-power potential}
\author {Yue Zhong \hspace {1cm}Lingshen Chen\hspace {1cm}Hongbo Cheng\footnote {E-mail address:
hbcheng@ecust.edu.cn}\\
Department of Physics, East China University of Science and
Technology,\\ Shanghai 200237, China}

\date{}
\maketitle

\begin{abstract}
We discuss the $U(1)$ gauged Q-balls with $N$-power potential to
examine their properties analytically. More numerical descriptions
and some analytical consideration have been contributed to the
models governed by four-power potential. We also demonstrate
strictly some new limitations that the stable $U(1)$ gauged
Q-balls should accept instead of estimating those with only some
specific values of model variables numerically. Having derived the
explicit expressions of radius, the Noether charge and energy of
the gauged Q-balls, we find that these models under the potential
of matter field with general power and the boundary conditions
will exist instead of dispersing and decaying. The Noether charge
of the large gauged Q-balls must be limited. The mass parameter of
the model can not be tiny.
\end{abstract}

\vspace{6cm} \hspace{1cm} PACS number(s): 98.80.-k, 95.35.+d,
11.27.+d

\newpage

\noindent \textbf{I.\hspace{0.4cm}Introduction}

There is considerable interest in the Q-balls as nontopological
solitons. The Q-balls are some kinds of nontopological systems
with defect structures possessing a conserved Noether charge
because of a symmetry of the Lagrangian [1-4]. The systems appear
in extended localized solution of models with certain
self-interacting complex scalar field having the minimum energy
configuration [1-4]. It was deduced that the Q-balls could form as
a part of solitogenesis during a phase transition in the early
universe and can survive now [5-10]. There are a lot of global
Q-balls showing that the matter or energy distributes within a
finite region. The spinning Q-balls are the flat spacetime limit
of rotating boson stars. A limited frequency range can allow the
spherically symmetric Q-balls and boson stars to exist [12, 13].
In those cases the coupling to gravity leads a spiral-like
frequency relating to the mass and charge of boson stars [12, 13].
The further discussion on the interacting boson stars and Q-balls
indicated that this kind of system can not be stable with
sufficiently large interaction [14]. The signum-Gordon typed
Q-balls put forward by H. Arodz et. al. are compact [15, 16] and
this kind of models in electrodynamics were also studied [17]. The
analytical discussions were presented on the Q-balls with
conical-shaped potential in the higher-dimensional spacetime [18].
Linear Q-balls with the description of a single complex scalar
field were constructed [19]. Q-balls can become the candidates of
dark matter because of their lifetime or cross section [20, 21].
The Q-balls were also be used to explain the origin of the baryon
asymmetry [22, 23] and so do the other kinds of the compact
objects [24]. In the case of gauge-mediated supersymmetric Q-balls
with large enough charge, the energy per unit charge is less than
the rest mass of one particle, so this kind of Q-balls are stable
instead of dispersing [6, 20]. Further this kind of Q-balls were
explored in the inflation scenario to explain the baryon asymmetry
and dark matter of the universe [25]. During the decay of the
gravity-mediated supersymmetric Q-balls, the lightest
supersymmetric particles can generate to become candidates of dark
matter and so do some other stable gauge-mediated-type Q-balls
[26-29]. The details of evolution of Q-ball were considered. In
the case of gauge-mediated SUSY breaking, the appearance of
unstable Q-balls is necessary for AD baryogenesis in GMSB in view
of the astrophysical constraints from the stability of neutron
stars. The Q-ball decay leads the gravitino dark matter [30]. Some
efforts were devoted to the Q-balls under the thermal logarithmic
potential showing the influence from the expansion of the universe
and this kind of Q-balls can also be used to describe the baryon
asymmetry and dark matter [31-34].

In addition to the global Q-balls mentioned above, the models of
gauged Q-balls also attract more attention. The nontopological
solitons with gauge $U(1)$ symmetry were studied [35]. Further the
so-called $U(1)$ gauged Q-balls were investigated [36]. A lot of
theoretical considerations were paid to the gauged Q-balls and the
valuable numerical estimations were obtained [17, 37-42].
Recently, I. E. Gulamov et.al. discussed the $U(1)$ gauged Q-balls
with the potential involving the highest power term as
$(\Phi^{+}\Phi)^{2}$ and compare their properties with those of
Q-balls in the nongauged case while the two kinds of Q-balls have
the scalar field potential in the same form to show the
considerable difference between the two types [43]. It was shown
that the Q-balls consisting of two kinds of complex scalar fields
and $U(1)$ gauge field survive in the universe, one complex scalar
field with positive electric charge having baryon charge, and the
other negative-electric-charged field carrying lepton charge [44].
The charged boson stars made of massive complex scalar fields
connecting the $U(1)$ gauge field and gravity governed by V-shaped
scalar potential were considered in the de Sitter or anti de
Sitter spacetime [45].

It is necessary to investigate the $U(1)$ gauged Q-balls under the
potential containing $(\Phi^{+}\Phi)^{\frac{N}{2}}$ term
analytically with the help of virial theorem. The formation and
characters of several kinds of gauged Q-balls have been evaluated
[43-45]. According to the analytical discussions and numerical
estimation, the important properties of $U(1)$ gauged Q-balls
subject to four-power potential were demonstrated [43]. M. Gleiser
et.al. put forward a generalized virial relation for Q-balls with
general potential in the spacetime with arbitrary dimensionality
to describe the Q-balls analytically instead of numerical
estimations for several values of model parameters shown as a
series of curves [46]. It is necessary to generalize the
consideration to the case with term $(\Phi^{+}\Phi)^{\frac{N}{2}}$
in the potential [45] for the Q-balls with interactions and
certainly their field equation is not easy to be discussed just in
view of the numerical calculation. We think that the reliable and
explicit relations among the model parameters consisting of the
gauged Q-ball are difficult to be exhibited by performing the
burden numerical calculation repeatedly because the field
equations for this kind of model are nonlinear. It is fundamental
to find the analytical expressions for Q-ball's charge, radius and
energy, the reliable and explicit relations among the model
parameters and gauged interaction for the existence of Q-balls.
The numerical results for a series of fixed values of system
parameters can not reveal the model properties and the influence
from interaction completely. The analytical description for gauged
Q-balls is more powerful. We are going to make use of these
results to scrutinize how the gauged interaction and the power $N$
affect the system feature. Here we plan to revisit the $U(1)$
gauged Q-balls with the $N$-power potential [43-45] analytically
in virtue of the scheme from Ref. [46] and hope to find the
necessary conditions that enable this kind of models to exist.
First we estimate the energy per unit charge with the help of
virial relation. We investigate the existence and stability in the
cases of large ball and small ones respectively.

This paper is organized as follow. We find the virial relation for
the $U(1)$ gauged Q-balls with $(\Phi^{+}\Phi)^{\frac{N}{2}}$
scalar field potential at first. Secondly we perform the
analytical discussion of the radius and energy of these Q-balls to
exhibit their properties in the cases of large radius or small
ones respectively. During the research on this kind of Q-balls, we
will find the upper limit to their Noether charge. The results are
listed in the end.

\vspace{0.8cm} \noindent \textbf{II.\hspace{0.4cm}The virial
relation for $U(1)$ gauged Q-balls with
$(\Phi^{+}\Phi)^{\frac{N}{2}}$ scalar field potential}

In order to explore the profile of the energy of $U(1)$ gauged
Q-balls, we employ the virial relation [46]. We choose the
Lagrangian density of the system consisting of scalar field with
$U(1)$ gauged symmetry as follows [43],

\begin{equation}
{\mathcal{L}}=(D_{\mu}\Phi)^{+}D^{\mu}\Phi-V(\Phi^{+}\Phi)
-\frac{1}{4}F_{\mu\nu}F^{\mu\nu}
\end{equation}

\noindent where

\begin{equation}
D_{\mu}=\partial_{\mu}+ieA_{\mu}
\end{equation}

\begin{equation}
F_{\mu\nu}=\partial_{\mu}A_{\nu}-\partial_{\nu}A_{\mu}
\end{equation}

\noindent and $\Phi=\Phi(x)$ is a complex scalar field, gauge
fields $A^{\mu}=A^{\mu}(x)$. The index $\mu=0, 1, 2, 3$ and the
signature is $(+, -, -, -)$. The field equations are [43],

\begin{equation}
D_{\mu}D^{\mu}\Phi+\frac{\partial}{\partial\Phi^{+}}V(\Phi^{+}\Phi)=0
\end{equation}

\begin{equation}
\partial_{\mu}F^{\mu\nu}=ie\Phi^{+}D^{\nu}\Phi
-ie(D^{\nu}\Phi)^{+}\Phi
\end{equation}

\noindent The potential governing the system has a generalized
form [43],

\begin{equation}
V(\Phi^{+}\Phi)=M^{2}\Phi^{+}\Phi-\lambda(\Phi^{+}\Phi)^{\frac{N}{2}}
\end{equation}

\noindent where $M$ is the mass of scalar field and the parameter
$\lambda$ is positive. It is obvious that the potential (6) has a
global minimum at $\Phi=0$ and admits the formation of Q-balls. In
order to become Q-balls, this kind of system with local $U(1)$
symmetry should carry a net particle number called $Q$ which is
conserved and should keep its energy to be smaller than $QM$ while
the complex scalar fields and gauge field satisfy some boundary
conditions to let the scalar fields distribute within a finite
region. The associated conserved current density is defined as
[2-4, 43-45],

\begin{equation}
j^{\mu}=-i[\Phi^{+}D^{\mu}\Phi-\Phi(D^{\mu}\Phi)^{+}]
\end{equation}

\noindent and the corresponding conserved charge can be given by
[2],

\begin{equation}
Q=\int j^{0}d^{3}x
\end{equation}

\noindent In the static case, the components of gauge field can be
chosen as [43],

\begin{equation}
A^{\mu}(x)=(A_{0}(x), 0, 0, 0)
\end{equation}

\noindent We take the ansatz for the fields configurations to be
spherically symmetry like [43],

\begin{equation}
\Phi(x)=f(r)e^{i\omega t}
\end{equation}

\begin{equation}
A_{0}({\mathbf{r}})= A_{0}(r)
\end{equation}

\noindent for the lowest energy. Here $r=|\mathbf{r}|$. There are
boundary conditions imposed on the fields describing the gauged
Q-balls [43],

\begin{equation}
\lim_{r\longrightarrow\infty}f(r)=0
\end{equation}

\begin{equation}
\lim_{r\longrightarrow\infty}A_{0}(r)=0
\end{equation}

\noindent According to Lagrangian (1), the total energy of the
system is [43],

\begin{eqnarray}
E=\int\mathcal{H}d^{3}x\hspace{7.5cm}\nonumber\\
=4\pi\int_{0}^{\infty}\left((\omega^{2}-e^{2}A_{0}^{2})f^{2}
+(\frac{\partial f}{\partial
r})^{2}+V(f)-\frac{1}{2}(\frac{\partial A_{0}}{\partial
r})^{2}\right)r^{2}dr
\end{eqnarray}

\noindent Here we hire the ansatz (10), (11) and

\begin{equation}
V(f)=M^{2}f^{2}-\lambda f^{N}
\end{equation}

\noindent The total Noether charge of the $U(1)$ gauged Q-balls
is,

\begin{eqnarray}
Q=\int j^{0}d^{3}x\hspace{2cm}\nonumber\\
=8\pi\int_{0}^{\infty}(\omega+eA_{0})f^{2}r^{2}dr
\end{eqnarray}

\noindent According to Ref. [46], the virial relation as a
generalization of Derrick's theorem for Q-balls can be expressed
as,

\begin{equation}
3<V(f^{2})>=\frac{3}{4<f^{2}>}[Q^{2}-4e^{2}(<A_{0}f^{2}>)^{2}]
+3e^{2}<A_{0}^{2}f^{2}>-<(\frac{\partial f}{\partial r})^{2}>
+\frac{1}{2}<(\frac{\partial A_{0}}{\partial r})^{2}>
\end{equation}

\noindent leading to,

\begin{eqnarray}
\frac{E}{Q}=\omega\{1+(1+\frac{2e}{Q}<A_{0}f^{2}>)
[1+\frac{6(<V(f^{2})>-e^{2}<A_{0}^{2}f^{2}>)}{2<(\frac{\partial
f}{\partial r})^{2}>-<(\frac{\partial A_{0}}{\partial
r})^{2}>}]^{-1}\}\nonumber\\
<M\hspace{11cm}
\end{eqnarray}

\noindent which keeps the Q-balls' stability. Here
$<\cdot\cdot\cdot>=\int\cdot\cdot\cdot d^{3}x$. If the ratio
$\frac{E}{Q}$ is small enough, the Q-balls will not decay into
several kinds of particles such as scalars, fermions etc. [20, 25,
29, 30]. It should be pointed out that the gauged Q-balls can
radiate because of the interactions among the scalar fields.

\vspace{0.8cm} \noindent \textbf{III.\hspace{0.4cm}The analytical
discussion on the large $U(1)$ gauged Q-balls with
$(\Phi^{+}\Phi)^{\frac{N}{2}}$ scalar field potential}

We are going to describe the $U(1)$ gauged Q-balls in the case of
huge Noether charge and radius and probe the necessary conditions
imposed on this kind of Q-balls as dark matter. As the first step
we follow the Coleman's procedure [1] to estimate the Q-balls. We
select the scalar field composing the Q-balls to be step functions
which are equal to be constants denoted as $f_{c}$ vanishing
outside the balls. For the static charged Q-balls with uniform
spherical charge distribution, the nonvanishing component of gauge
field can be chosen as $A_{0}(r)=\frac{eQ}{8\pi
R}(3-\frac{r^{2}}{R^{2}})$ and $A_{0}(r)=\frac{eQ}{4\pi r}$ within
the Q-ball and outside ones respectively and $R$ is Q-ball's
radius [47]. This choice satidfies the Eq. (13) and conditions
$\frac{dA_{0}(r)}{dr}|_{r=0}=0$ from Ref. [43]. The system energy
is,

\begin{equation}
E=\frac{3Q^{2}}{16\pi R^{3}f_{c}^{2}}-\frac{9Q^{2}e^{2}}{20\pi
R}-\frac{Q^{2}e^{4}f_{c}^{2}R}{700\pi}+\frac{4}{3}\pi
R^{3}V(f_{c}^{2})
\end{equation}

\noindent where $V(f_{c}^{2})=M^{2}f_{c}^{2}-\lambda f_{c}^{N}$.
We estimate the energy to find that the function can keep
positive.

Now we follow the procedure from Ref. [46] to introduce the
following field profiles to demonstrate the true large $U(1)$
gauged Q-balls,

\begin{eqnarray}
f(r)=\{\begin{array}{cc}
  f_{c} & r<R \\
  f_{c}e^{-\alpha(r-R)} & r\geq R \\
\end{array}
\end{eqnarray}

\noindent where $\alpha$ is positive variational parameters and
$R$ represents a region where the fields of Q-balls distribute.
Within the region this kind of field configuration keeps constant
instead of diminishing quickly, which means that the scalar field
of Q-balls can extend a little widely. According to the
large-Q-ball ansatzs (20), the conserved charge is,

\begin{eqnarray}
Q=\int j^{0}d^{3}x\hspace{9.5cm}\nonumber\\
=16\pi\omega f_{c}^{2}[\frac{1}{(2\alpha)^{3}}
+\frac{R}{(2\alpha)^{2}}+\frac{R^{2}}{4\alpha}+\frac{R^{3}}{6}]
+8e^{2}Qf_{c}^{2}[\frac{1}{4}\frac{1}{(2\alpha)^{2}}
+\frac{1}{4}\frac{R}{2\alpha}+\frac{R^{2}}{10}]
\end{eqnarray}

\noindent the relation between the topological charge $Q$ and the
frequency $\omega$. With substituting the field profile (20) and
the gauge field $A_{0}(r)$ mentioned before into the total energy
(14), the dominant part of energy of the gauged Q-ball reads,

\begin{eqnarray}
E=E[f, A_{0}]\hspace{2cm}\nonumber\\
\approx\frac{\alpha_{Q}^{2}}{R^{3}}+\beta_{Q}^{2}+a_{l}R^{2}
+b_{l}R^{3}
\end{eqnarray}

\noindent where

\begin{equation}
\alpha_{Q}^{2}=\frac{3Q^{2}}{16\pi f_{c}^{2}}
\end{equation}

\begin{equation}
\beta_{Q}^{2}=\frac{3Q^{2}e^{4}f_{c}^{2}}{25\pi}
\end{equation}

\begin{equation}
a_{l}=2\pi \alpha f_{c}^{2}+\frac{2\pi}{\alpha}(M^{2}f_{c}^{2}
-\frac{2\lambda f_{c}^{N}}{N})
\end{equation}

\begin{equation}
b_{l}=\frac{4\pi}{3}(M^{2}f_{c}^{2}-\lambda f_{c}^{N})
\end{equation}

\noindent The above approximation is performed by leaving several
dominant terms in the expression of the energy and this
approximation is acceptable for large Q-balls.

In order to investigate the stability of this kind of Q-balls and
the constrains on them, we extremize the total energy with respect
to $R$ to determine the minimum energy. We proceed the performance
$\frac{\partial E}{\partial R}\mid_{R=R_{cl}}=0$ to find the
equation that the critical radius $R_{cl}$ of Q-balls refers to,

\begin{equation}
3b_{l}R_{cl}^{6}+2a_{l}R_{cl}^{5}+\beta_{Q}^{2}R_{cl}^{4}
-3\alpha_{Q}^{2}=0
\end{equation}

\noindent The approximate solution to Eq.(27) is,

\begin{equation}
R_{cl}=(\frac{3}{32\pi f_{c}^{2}b_{l}})^{\frac{1}{6}}\xi
Q^{\frac{1}{3}}-\frac{a_{l}}{9b_{l}+\frac{6e^{4}f_{c}^{2}}{25\pi\xi^{2}}
(\frac{32\pi f_{c}^{2}b_{l}}{3})^{\frac{1}{3}}Q^{\frac{4}{3}}}
\end{equation}

\noindent where

\begin{equation}
\xi^{2}=[(1-\frac{2\beta_{Q}^{6}}{729\alpha_{Q}^{2}b_{l}^{2}})
+\sqrt{1-\frac{4\beta_{Q}^{6}}{729\alpha_{Q}^{2}b_{l}^{2}}}]^{\frac{1}{3}}
+[(1-\frac{2\beta_{Q}^{6}}{729\alpha_{Q}^{2}b_{l}^{2}})
-\sqrt{1-\frac{4\beta_{Q}^{6}}{729\alpha_{Q}^{2}b_{l}^{2}}}]^{\frac{1}{3}}
-\frac{2^{\frac{1}{3}}\beta_{Q}^{2}}{9(\alpha_{Q}b_{l})^{\frac{2}{3}}}
\end{equation}

\noindent We find that the minus term involving mass parameter
will become smaller with larger $M$. In order to keep the critical
radius of the Q-ball real, the root should not be negative,

\begin{equation}
1-\frac{4\beta_{Q}^{6}}{729\alpha_{Q}^{2}b_{l}^{2}}\geq0
\end{equation}

\noindent leading,

\begin{equation}
Q\leq\sqrt{\frac{375}{2}}\pi\frac{\sqrt{\mid M^{2}-\lambda
f_{c}^{N-2}\mid}}{e^{3}f_{c}}
\end{equation}

\noindent We find the upper limit on the Noether charge of the
gauged Q-ball, which is also consist with the conclusion drawn
with numerical estimation in the case of four-power field
potential [43]. Our results involving the power $N$ and the gauge
coupling indicate that the Noether charge must be limited or the
Q-ball radius will not be real. The expression (31) declares that
the upper limit will approach the infinity if the gauge coupling
$e$ vanishes, which means that no restriction on the charge $Q$ in
the global case.

Further we impose the condition $\frac{\partial E}{\partial
\alpha}\mid_{\alpha=\alpha_{c}}=0$ into Eq. (28) to obtain,

\begin{equation}
\alpha_{c}^{2}=M^{2}-\frac{2\lambda f_{c}^{N-2}}{N}
\end{equation}

\noindent which requires that $M>(\frac{2}{N}\lambda
f_{c}^{N-2})^{\frac{1}{2}}$ similar to that of Ref. [43]. The
energy of the $U(1)$ gauged Q-ball with the critical radius and
critical parameter is,

\begin{equation}
\frac{E[f]\mid_{R=R_{cl}, \alpha=\alpha_{c}}}{Q}
=4b_{l}(\frac{3}{32\pi f_{c}^{2}b_{l}})^{\frac{1}{2}}
(\frac{2}{\xi^{3}}+\xi^{3})+4\pi
f_{c}^{2}\alpha_{c}(\frac{3}{32\pi f_{c}^{2}b_{l}})^{\frac{1}{3}}
\xi^{2}Q^{-\frac{1}{3}}
\end{equation}

\noindent The asymptotic behaviour of the energy per unit charge
of large charged Q-balls with critical radius $R_{cl}$, critical
parameter $\alpha_{c}$ and huge charge $Q$ is,

\begin{equation}
\lim_{Q\longrightarrow\infty}\frac{E[f]\mid_{R=R_{cl},
\alpha=\alpha_{c}}}{Q}=4b_{l}(\frac{3}{32\pi
f_{c}^{2}b_{l}})^{\frac{1}{2}}(\frac{2}{\xi^{3}}+\xi^{3})
\end{equation}

\noindent It is obvious that the above explicit expression is
finite and its result can be regulated to be lower than the
kinetic energy. The energy per unit charge lower than the kinetic
energy per unit charge can keep this kind of $U(1)$ gauged Q-balls
to survive instead of dispersing. The Figure 1 shows the
dependence of the minimum energy per unit charge of $U(1)$ gauged
Q-balls with $N$-power potential on charge $Q$ for power $N$ with
$f_{c}=2$ for simplicity. It is obvious that the stronger
potential involving higher power increases the energy density. We
can select the magnitudes of model parameters to keep the energy
density lower than the necessary kinetic energy of a free particle
in order to lead the Q-balls stable. For various values of power
$N$, the shapes of curves of minimum energy over charge are
similar.

\vspace{0.8cm} \noindent \textbf{IV.\hspace{0.4cm}The analytical
discussion on the small $U(1)$ gauged Q-balls with
$(\Phi^{+}\Phi)^{\frac{N}{2}}$ scalar field potential}

Here we focus on the small $U(1)$ gauged Q-balls with the help of
scheme from Ref. [46] and discuss the possibility that this kind
of small balls could be dark matter. The small $U(1)$ gauged
Q-balls that we will consider is also controlled by the potential
(15) from Eq.(6) [43]. We bring about the Gaussian ansatz like
[46],

\begin{equation}
f(r)=f_{c}e^{-\frac{r^{2}}{R^{2}}}
\end{equation}

\noindent where $f_{c}$ is constant. This kind of function lets
the field decrease directly and rapidly and certainly the size of
the Q-ball will be small. Substituting the ansatz (35) into the
expression (16), we write the conserved charge belonging to the
small gauged Q-balls as follows,

\begin{equation}
Q=\sqrt{\frac{\pi^{3}}{2}}\omega
f_{c}^{2}R^{3}+e^{2}Qf_{c}^{2}R^{2}[\frac{3}{16e^{2}}+\frac{9}{32}
\sqrt{\frac{\pi}{2}}\textmd{erf}(\sqrt{2})]
\end{equation}

\noindent where $\textmd{erf}(z)$ is the error function [47]. The
total energy in the case of small Q-balls is,

\begin{eqnarray}
E=E[f_{c}]\hspace{2.5cm}\nonumber\\
\approx\frac{a_{Q}}{R^{3}}+b_{s}R+c_{s}R^{3}+\frac{d_{s}}{R}
\end{eqnarray}

\noindent where

\begin{equation}
a_{Q}=\frac{Q^{2}}{\sqrt{2}\pi^{\frac{3}{2}}f_{c}^{2}}
\end{equation}

\begin{equation}
b_{s}=\frac{e^{4}Q^{2}f_{c}^{2}}{\sqrt{2}\pi^{\frac{3}{2}}}
[\frac{3}{16e^{2}}+\frac{9}{32}\sqrt{\frac{\pi}{2}}\textmd{erf}(\sqrt{2})]^{2}
-\frac{e^{4}Q^{2}f_{c}^{2}}{16\pi}[\sqrt{2\pi}-\frac{27}{64e^{2}}
-\frac{169}{128}\sqrt{\frac{\pi}{2}}\textmd{erf}(\sqrt{2})]
+3(\frac{\pi}{2})^{\frac{3}{2}}f_{c}^{2}
\end{equation}

\begin{equation}
c_{s}=(\frac{\pi}{2})^{\frac{3}{2}}[M^{2}f_{c}^{2}
-(\frac{2}{N})^{\frac{3}{2}}\lambda f_{c}^{N}]
\end{equation}

\begin{equation}
d_{s}=\{-\frac{\sqrt{2}}{\pi^{\frac{3}{2}}}[\frac{3}{16e^{2}}
+\frac{9}{32}\sqrt{\frac{\pi}{2}}\textmd{erf}(\sqrt{2})]
-\frac{3}{20\pi}\}e^{2}Q^{2}
\end{equation}

\noindent In order to establish the equation for the critical
radius $R_{cs}$, we extremized the expression of the energy with
respect to $R$ like $\frac{\partial E}{\partial
R}\mid_{R=R_{cs}}=0$, then

\begin{equation}
3c_{s}R_{cs}^{6}+b_{s}R_{cs}^{4}-h_{s}R_{cs}-3a_{Q}=0
\end{equation}

\noindent showing the acceptable approximate solution as,

\begin{equation}
R_{cs}\approx
R_{0}[1-\frac{1}{18c_{s}R_{0}^{2}}(b_{s}-\frac{h_{s}}{R_{0}^{2}})]
\end{equation}

\noindent where

\begin{equation}
R_{0}=\{\frac{2Q^{2}}{\pi^{3}f_{c}^{2}[M^{2}f_{c}^{2}-
(\frac{2}{N})^{\frac{3}{2}}\lambda f_{c}^{N}]}\}^{\frac{1}{6}}
\end{equation}

\noindent where the magnitude must be real with
$M>[(\frac{2}{N})^{\frac{3}{2}}\lambda f_{c}^{N-2}]^{\frac{1}{2}}$
similar to that from Ref. [43]. It is obvious that this magnitude
is smaller than the previous requirement $(\frac{2}{N}\lambda
f_{c}^{N-2})^{\frac{1}{2}}$ in the case of larger balls. In a word
the parameter $M$ can not be too small. The minimum energy per
unit charge for small $U(1)$ gauged Q-balls can be expressed in
terms of the critical radius as,

\begin{eqnarray}
\frac{E[f_{c}]\mid_{R=R_{cs}}}{Q}
=2c_{s}(\frac{1}{\sqrt{2}\pi^{\frac{3}{2}}}
f_{c}^{2}c_{s})^{\frac{1}{2}}[1+\frac{b_{s}}{2c_{s}}(\sqrt{2}\pi^{\frac{3}{2}}
f_{c}^{2}c_{s})^{\frac{1}{3}}Q^{-\frac{2}{3}}\hspace{3cm}\nonumber\\
-(\frac{b_{s}^{2}}{36c_{s}^{2}}
-\frac{d_{s}}{2c_{s}})(\sqrt{2}\pi^{\frac{3}{2}}f_{c}^{2}c_{s})^{\frac{2}{3}}
Q^{-\frac{4}{3}}+\frac{b_{s}d_{s}}{36c_{s}^{2}}(\sqrt{2}\pi^{\frac{3}{2}}f_{c}^{2}c_{s})
Q^{-2}]
\end{eqnarray}

\noindent In fact the smaller Q-balls can also swallow more
charge, so the minimum energy per unit charge is exhibited as,

\begin{equation}
\lim_{Q\longrightarrow\infty}\frac{E[f_{c}]}{Q}\mid_{R=R_{cs}}
=2c_{s}(\frac{1}{\sqrt{2}\pi^{\frac{3}{2}}}
f_{c}^{2}c_{s})^{\frac{1}{2}}
\end{equation}

\noindent The above magnitude is finite and can also be regulated
to be small for a series of special values of model parameters to
keep the small $U(1)$ gauged Q-balls stability instead of
dispersing and decaying. In Figure 2, for comparison we also
select $f_{c}=2$ without losing generality and show that the
functions of minimum energy per unit charge of small $U(1)$ gauged
Q-balls on charge $Q$ and power $N$ are similar to those of large
ones. The smaller Q-balls can survive under the influence from
$U(1)$ interaction.

\vspace{0.8cm} \noindent \textbf{IV.\hspace{0.4cm}Summary and
Conclusion}

Here we study the $U(1)$ gauged Q-balls with $N$-power field
potential strictly by means of variational estimation without
solving the nonlinear field equations numerically with respect to
several given values of model parameters. In particular, the
analytical expressions for radius and energy of this kind of
Q-ball are obtained. Our analytical discussions help us to declare
that the energy per unit charge of Q-balls made of charged scalar
field will not be divergent if the charge is large. The ratio of
total energy and charge can be sufficiently small for the ranges
of special values imposed on the Q-balls construction to keep
their stability instead of dispersing or decaying, which are
certainly consistent with results in the case of $N=4$ [43]. Our
analytical expressions are explicit and reliable. The radius and
energy of Q-ball have the terms associated with the gauge coupling
$e$ whose effect on the Q-balls' properties is considerable and
certainly the characters of gauged Q-balls are different from
those of global ones. It should be indicated explicitly that the
mass parameter can not be too small like $M>(\frac{2}{N}\lambda
f_{c}^{N-2})^{\frac{1}{2}}$ based on our explicit expression for
Q-ball radius. The Noether charge for the potential with $N=4$ is
within a region which was estimated with series of numerical
consideration in Ref. [43]. With the help of scheme [46], we
discover the radii of the $U(1)$ gauged Q-balls analytically and
keep the balls radii to be real to obtain the explicit expression
for charge's upper limit associated with the model variable, the
gauge coupling and the highest power $N$ when the gauged Q-balls
under the general potential are rather large. Some further works
have been proceeded.

\vspace{3cm}

\noindent\textbf{Acknowledgement}

This work is supported by NSFC No. 10875043.

\newpage
\begin{figure}
\setlength{\belowcaptionskip}{10pt} \centering
\includegraphics[width=15cm]{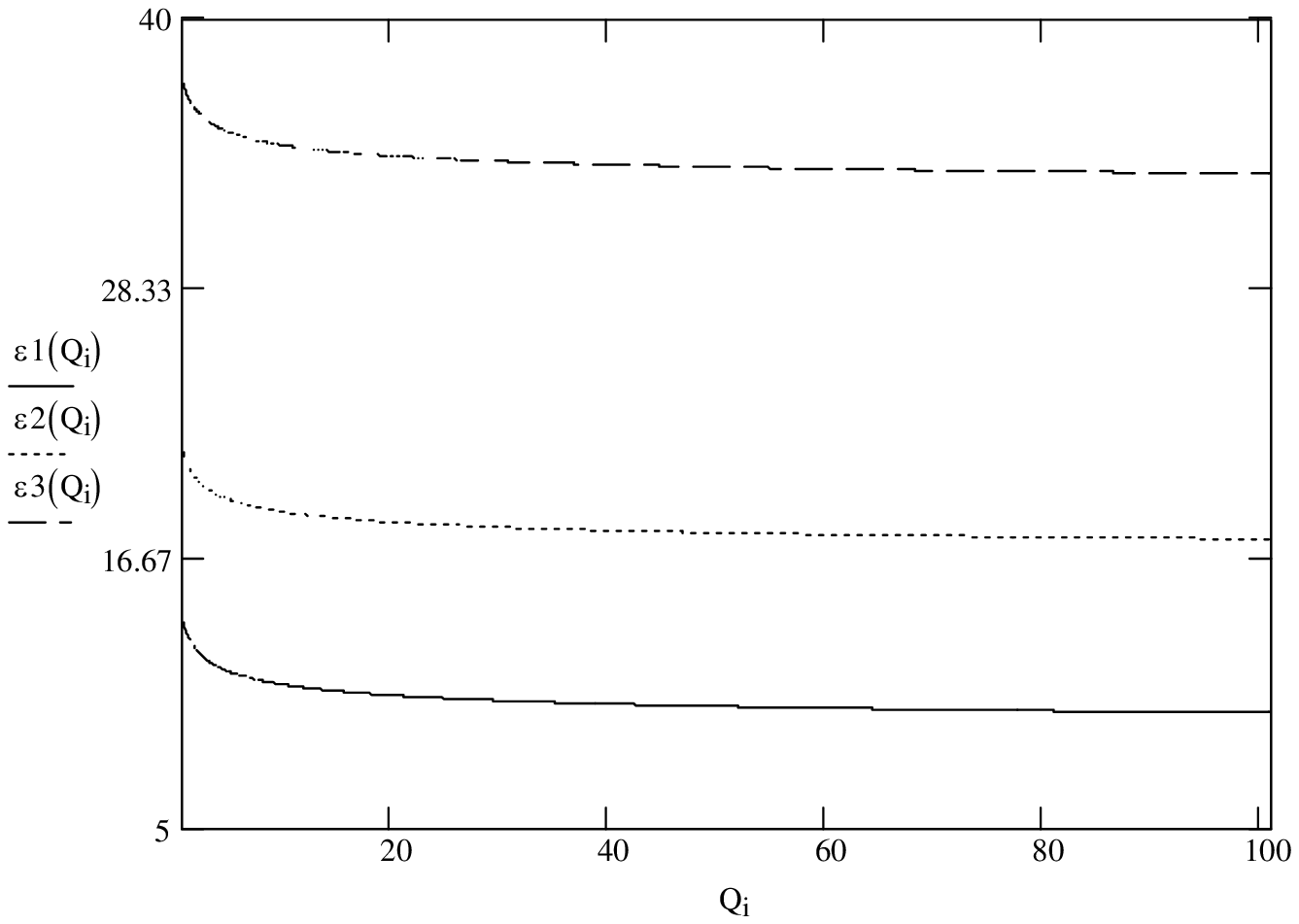}
\caption{The solid, dot, dashed curves of the minimum energy per
unit charge of large $U(1)$ gauged Q-balls in the $N$-power
potential as functions of charge $Q$ for power $N=4, 6, 8$
respectively.}
\end{figure}

\newpage
\begin{figure}
\setlength{\belowcaptionskip}{10pt} \centering
\includegraphics[width=15cm]{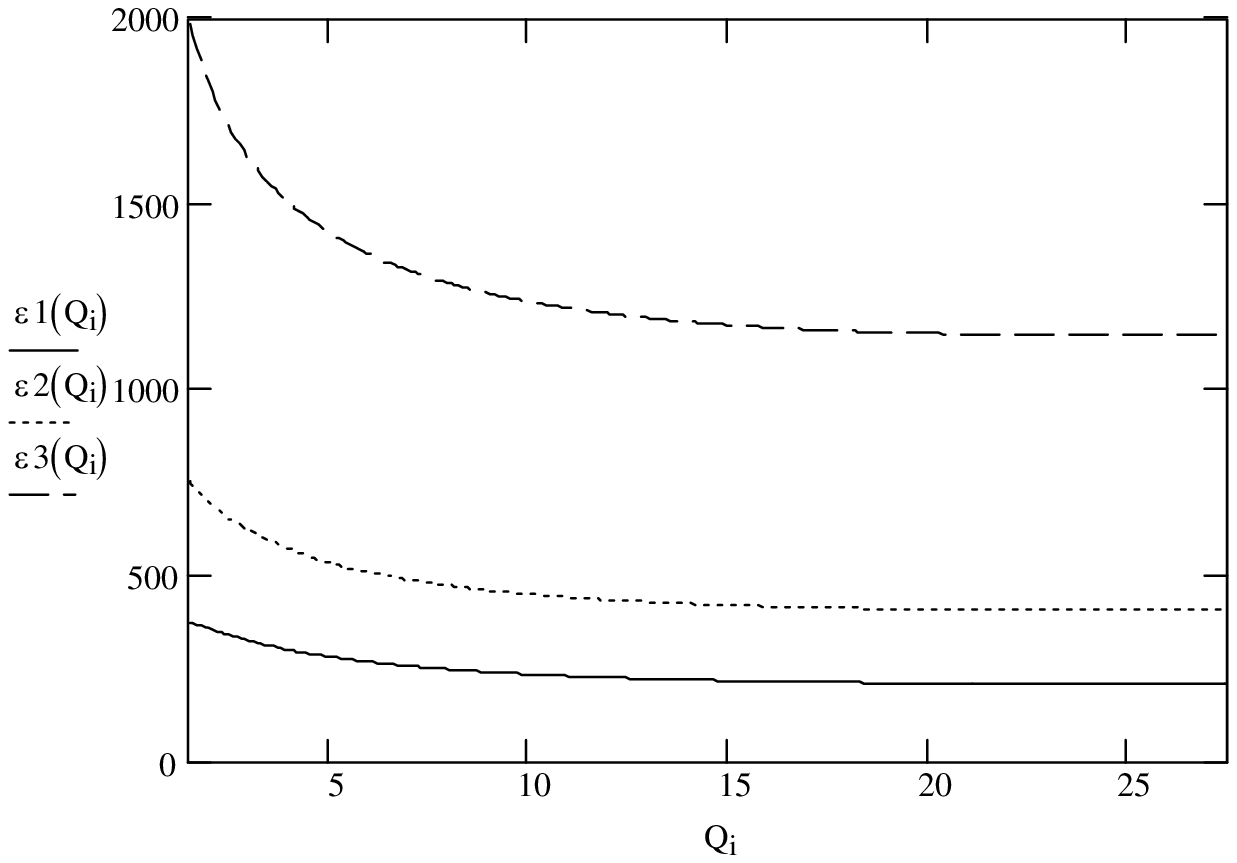}
\caption{The solid, dot, dashed curves of the minimum energy per
unit charge of small $U(1)$ gauged Q-balls in the $N$-power
potential as functions of charge $Q$ for power $N=4, 6, 8$
respectively.}
\end{figure}

\end{document}